\documentclass[twocolumn]{aastex631}

\usepackage{amsmath}
\usepackage{tabularx}
\usepackage{mathrsfs}
\usepackage{xcolor}
\usepackage{CJK}
\usepackage{tablefootnote}

\newcommand\free{\text{free}}
\newcommand\forced{\text{forced}}

\shorttitle{Free inclinations of TNOs}
\shortauthors{Huang et al.}

\graphicspath{{./}{figures/}}

\begin{document}

\begin{CJK*}{UTF8}{gbsn}
\title{Free Inclinations for Transneptunian Objects in the Main Kuiper Belt}

\author[0000-0003-1215-4130]{Yukun Huang (黄宇坤)}

\author[0000-0002-0283-2260]{Brett Gladman}
\affiliation{Dept. of Physics and Astronomy \\
University of British Columbia \\
6224 Agricultural Road \\
Vancouver, BC V6T 1Z1, CANADA}

\author[0000-0001-8736-236X]{Kathryn Volk}
\affiliation{Lunar and Planetary Laboratory \\
1629 E University Blvd \\
Tucson, AZ, 85721, USA}

\begin{abstract}
There is a complex inclination structure present in the transneptunian
object (TNO) orbital distribution in the main classical belt region
(between orbital semimajor axes of 39 and 48 au).
The long-term gravitational effects of the giant planets
make TNO orbits precess, but non-resonant objects maintain a
nearly constant `free' inclination ($I_\free$) with respect to a local forced precession pole.
Because of the likely cosmogonic importance of the distribution of
this quantity, we tabulate free inclinations for all main-belt TNOs,
each individually computed using barycentric orbital elements with
respect to each object's local forcing pole.
We show that the simplest method, based on the Laplace-Lagrange secular theory, is unable to give correct forcing poles for objects near the $\nu_{18}$ secular resonance, resulting in poorly conserved $I_\free$ values in much of the main belt.
We thus instead implemented an averaged Hamiltonian to obtain the expected nodal precession for each TNO, yielding significantly more accurate free inclinations for non-resonant objects. For the vast majority (96\%) of classical belt TNOs, these $I_\free$ values are conserved to $<1^\circ$ over 4 Gyr numerical simulations,
demonstrating the advantage of using this well-conserved quantity in studies of the TNO population and its primordial inclination profile; our computed distributions only reinforce the idea of a very co-planar surviving `cold' primordial population, overlain by a large $I$-width implanted `hot' population.
\end{abstract}

\keywords{Trans-Neptunian objects (1705) ---  Kuiper belt (893)--- Celestial mechanics (211)}

\section{Introduction} \label{sec:intro}
\end{CJK*}
The outer region of our Solar System beyond Neptune (transneptunian space) hosts a large swarm of icy bodies that are planetesimals left over after the planet formation era. They contain valuable information about the Solar System's distant past. Over the past two decades, there is growing consensus that current transneptunian objects (TNOs) may have accreted from different regions in the protoplanetary disk: the cold population likely formed locally around $a\approx44$ au and hasn't experienced significant subsequent dynamical excitation or collisional evolution, whereas the hot population likely formed closer to the Sun ($a < 30$ au) and was implanted at the current locations during the late stages of planet formation.
These two populations are mixed in the main Kuiper Belt ($42 \lesssim a \lesssim 47$ au) radially and vertically, with the inclination being a rough proxy to separate them (see \citealt{Gladman.2021} for a detailed review).

\citet{Brown.2001} first fit the inclination distribution of all classical TNOs (i.e. main belt TNOs not in mean motion resonances with Neptune) using a functional form of $\sin I$ multiplied by a sum of two Gaussians consisting of a cold component (of width $\sim 2.2^\circ$) and a hot component ($\sim17^\circ$). \citet{Levison.2001} noticed the observed cold population lacks large objects, which was later further confirmed by several independent Kuiper belt surveys showing the cold population has a significantly steeper absolute magnitude (H) distribution than the hot population \citep{Bernstein.2004, Elliot.2005, Fraser.2010, Petit.2011, Kavelaars.2021}. 
The perihelion distance distribution of the cold population is more confined than those of the hot \citep{Petit.2011}. 
The cold classicals are also known to have a higher abundance of binary TNOs \citep[see, e.g.][]{Noll.2020}, especially those with comparable sizes. Furthermore, a statistically significant correlation between the color and inclination of the classical objects has been observed, with low-inclination objects more likely to be red and high-inclination objects likely to be more neutral in color \citep{Doressoundiram.2002, Trujillo.2002, doressoundiram2008color, Peixinho.2008}. High-precision colors from optical and near-infrared observations have demonstrated that the cold classicals have different surface properties than the hot members \citep{Pike2017, Schwamb.2019, Muller.2020, 10.3847/psj/abc34e}. All of these properties are consistent with the two populations having distinct formation histories. As a result, the orbital distributions, especially the inclination distributions, of the two populations shed light upon their dynamical past and deserve detailed investigation with the most recent sample.

A common practice in TNO research is to split the classical TNOs into hot and cold populations with a simple inclination cut 
to facilitate, for example, physical property studies of the two populations or comparisons between observationally derived population estimates and those from dynamical models.

For example, \citet{Bernstein.2004}, \citet{Petit.2011} and \citet{Fraser.2014} all used a  cut of $I<5^\circ$ in ecliptic inclination
to identify a dominantly cold population, 
while \citet{Peixinho.2008} used $12^\circ$.
The ecliptic $I$, however, varies over time as an artifact of the reference frame choice: a TNO's orbit precesses around its local forcing pole with a fixed $I_\free$ and a constant frequency, the result of which, in ecliptic space, is a non-constant precession (sometimes not even a precession but a confined oscillation in the longitude of ascending node $\Omega$) with a varying $I$ (see section 7.4 of \citealt{Murray.1999} or figure 1 of \citealt{Gladman.2021}). 
This naturally makes $I_\free$, a conserved quantity regardless of the choice of reference frame, preferable to the ecliptic inclination, which is the commonly tabulated quantity.

We note that because the real classical belt TNO distribution is a sum of two overlapping components that each have different inclination widths,
there will always be some level of contamination when using a simple cut (see fig.2 of \citet{Dawson.2012}, for example).
Cutting in $I_\free$ rather than ecliptic $I$, however,  dramatically improves how well the two components are isolated.

\citet{Laerhoven.2019} showed that when using a free inclination cut of $4^\circ$, the cold classical TNOs are best fit with a narrower width of $\simeq1.75^\circ$, strongly limiting its past perturbation. \citet{Gladman.2021} also found this cut results in a cleaner separation in TNO colors (their figure 6). 
Because of this superiority, the $4^\circ$ cut in free inclination is also adopted in \citet{Kavelaars.2021}.

With today's large TNO sample, including survey data from the Canada-France Ecliptic Plane Survey (CFEPS, \citealt{Jones.2006}), 
the Deep Ecliptic Survey \citep{Adams.2014},
the Outer Solar System Origins Survey (OSSOS, \citealt{Bannister.2018}), and the Dark Energy Survey \citep{10.3847/1538-4365/ac3914}, 
it is thus necessary to compute $I_\free$ for each main belt TNO. 
To do this, the local forcing planes or the forcing poles, relative to which $I_\free$ is measured, must be correctly calculated.
\citet{Brown.20041og} first realized the apparent mean plane of the TNOs differs significantly from the solar system's invariable plane (the plane defined by the average angular momentum of the larger planets).
In contrast, \citet{Elliot.2005} found the mean plane of the classical TNOs is more consistent with the invariable plane than with the local Laplacian plane (the latter being the plane perpendicular to the local forcing pole discussed below).
In a subsequent study, \citet{Chiang.2008} investigated the theoretically predicted locations of forcing poles, pointing out that the classical belt plane is significantly warped by the $\nu_{18}$ secular resonance near $a=40.5$ au; i.e., the local forcing plane in the main belt changes significantly with semimajor axis.
They also confirmed the conservation of TNO free inclination with respect to their calculated time-variable poles for 4 Gyr, but only for objects away from the singularity associated with the secular resonance.

Given that the calculation of the forcing poles (and thus the free inclinations) is somewhat complicated near secular resonances and that a non-negligible portion of the classical belt is affected by this, a better approach to calculating free inclinations is warranted.
In the present work, we implement a new method based on doubly averaging the Hamiltonian to obtain the expected nodal precession rates and the correct forcing pole for each TNO.
The free inclinations generated by this new algorithm represent a significant improvement over those given by the often-used linear secular theory, especially for objects 
within a few au of
the $\nu_{18}$ secular resonance singularity (see Section~\ref{sec2.2}). We thus tabulated the correct $I_\free$ of each main-belt TNO along with its barycentric orbital elements in Table~\ref{tab:free}.

\section{Computation of Free Inclinations}\label{sec2}

Because of the cosmogonic significance of the cold and hot populations, both for the dynamical structure of the transneptunian region and the interpretation of surface properties inferred from photometry and spectra, we chose to compute and publish TNO free inclinations.
Because as a {\it population} the cold objects exist only in the main belt between the 3/2 and 2/1 mean motion resonances with Neptune, our interpretation is that this component split is only sensible in this semimajor axis range (objects that might have low inclinations at other semimajor axes are best interpreted as the low-$I$ tail of the implanted hot component's inclination distribution); we thus confine ourselves to the main belt objects in this work.
In Section~\ref{sec2.1}, we describe how we selected the observed TNOs to include in our analysis.
Section~\ref{sec2.2} describes our approach to calculating free inclinations (with more details given in Appendix~\ref{appendix}), and Section~\ref{sec2.3} demonstrates that the newly calculated free inclinations are a better-conserved quantity than those calculated using linear theory.

\subsection{Dynamical Classification of TNOs}\label{sec2.1}

We began by downloading the most recent sample of main belt TNOs from the JPL Small-Body Database\footnote{\url{https://ssd.jpl.nasa.gov/sbdb_query.cgi}, retrieved on October 5th, 2021.}. We constrained the \textit{heliocentric} semimajor axis $a$ to the range of $(39.4, 47.7)$ au and the 1-sigma uncertainty in $a$ to $\delta a<0.1$ au. We also added two additional TNOs to the sample: 486958 Arrokoth (2014 MU$_{69}$), the target TNO visited by the New Horizons spacecraft on Jan. 1, 2019 \citep{Stern.2019} whose orbit-fit accuracy is not accurately reflected in JPL's database, and 2005 JY$_{185}$, an OSSOS object whose
$a$ uncertainty given by JPL is slightly above the $0.1$ au but our own estimate
is $\delta a \simeq 0.05$ au \citep{Bannister.2018}.
This resulted in 2018 objects being selected.

The JPL Small-Body Database provides orbital elements in the heliocentric IAU76/J2000 ecliptic reference frame. However, for the study of TNO dynamics, barycentric orbital elements are preferable to heliocentric ones. TNOs are relatively far away from the Sun, and the small wobbling of the Sun's position under planetary perturbations (mainly from the giant planets) leads to variation of TNO heliocentric orbital elements on timescales of each giant planet's orbital period \citep{Gladman.2021}. The barycenter of the Solar System, on the other hand, is far more stable in the eyes of TNOs. As a result, orbital elements in this paper and the appended table are all barycentric elements.

These TNOs are distributed across the whole main classical belt, starting from the 3/2 neptunian mean-motion resonance ($a =$ 39.4 au) and ending at the 2/1 resonance (47.7 au).
Several low-order resonances are also embedded in the main belt, the most important of which are the 5/3 (42.3 au), the 7/4 (43.7 au), and the 9/5 (44.6 au).
We classified the 2018 TNOs in our sample according to their current dynamical state \citep{Gladman.2008}, separating the classical TNOs from the resonant objects and the scattering objects (note that by definition detached objects have $a > $ 47.7 au, therefore they cannot be in this sample). 

To do this, we integrated the best-fit orbit for each TNO forward 10 Myr in time under the influence of the Sun and the four giant planets.
We used the \textsc{mercurius} algorithm within the \textsc{rebound} orbital integration software package \citep{Rein.2012}; this algorithm uses \textsc{rebound}'s \textsc{whfast} symplectic integrator \citep{whfast} for the majority of time steps and the adaptive-stepsize \textsc{ias15} integrator \citep{ias15} to resolve close encounters between test particles and planets.
We used a base time step of 0.25 years and an output interval of 1,000 years for these integrations.

After the integration, the TNO classification was carried out manually: each particle's $a$, eccentricity ($e$), and critical angle for the closest resonance ($\varphi$) are plotted. 
A human operator then decided to tag it as 
\textit{scattering} ($a$ alters more than 1 au), \textit{resonant} ($\varphi$ ceases to circulate at any moment in the 10 Myr integration), 
or \textit{classical} (for a non-scattering and non-resonant particle). 
Although recent papers have described TNO classification using automatic pipelines \citep{Khain.2020} or machine learning algorithms \citep{Smullen.2020}, we decided to do the job manually as the sample was not too large and this remains the most accurate method.
Our criterion for resonant objects is quite loose; this is motivated by the fact that even a brief interaction with a mean-motion
resonance can significantly alter $I_\free$ (see Section~\ref{sec2.3}).
For the resonant identification, we searched through resonances $i:j$ with $i = 1\ldots20$ and $j = 1\ldots20$ in the $a$ range of $(39.4, 47.7)$ au, which includes 23 distinct resonances with the 20:11 being the highest-order one.

Among the sample of 2018 objects we integrated and classified, 66\% (1332/2018) are classical, 31\% (622/2018) are resonant, and only 3\% (64/2018) are scattering.
These percentages have biases and should not be interpreted as the intrinsic or cosmogonic
dynamical distribution in this semimajor axis range\footnote{In particular, many of the scattering objects in this semimajor axis range have perihelia well inside of Neptune and were only detected with their faint absolute magnitudes because of their small current heliocentric distances; they are thus over-represented in our sample compared to the classical and resonant objects with brighter absolute magnitudes.}.
Note that our TNO classification is conducted with the purpose of better presenting our results in the next section; it should not be treated as the `definitive' classification for these objects because we are not considering orbital uncertainties by integrating clone orbits.

\subsection{Free Inclination with the Correct Precession Rate}\label{sec2.2}

\begin{figure*}[htb!]
  \centering
  \includegraphics[width=0.8\textwidth]{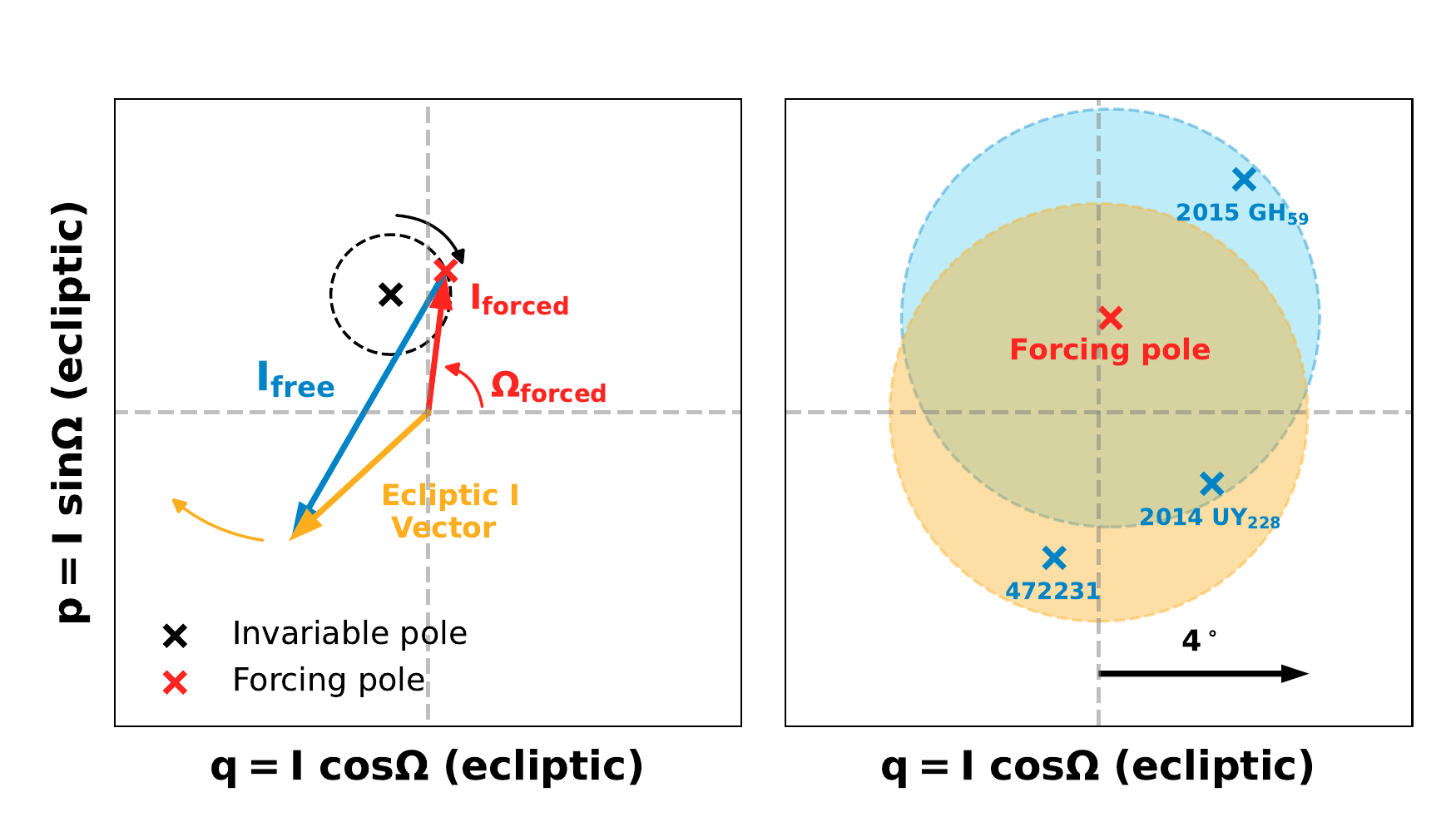}
  \caption{\textbf{Left panel:} Polar and rectangular coordinate depiction of the ecliptic (orange), the forced (red), and the 
  resultant free (blue) inclination vectors of a TNO in the main classical belt.
  For this object, the forcing center (red cross) rotates around the solar system's invariable pole (black cross) with a $\simeq$2 Myr period, the path of which is the black dashed circle.
  The free inclination (blue) vector rotates around the time-varying forcing center at a constant nodal precession rate $B$, keeping its magnitude $I_\free$ unchanged.
  The composition of these two movements gives rise to a more complicated inclination evolution in the ecliptic reference frame (orange vector with arrow denoting its approximate sense of motion).
  \textbf{Right panel:} The difference between a $4^\circ$ cut in ecliptic inclination (orange circle centered at origin) and a $4^\circ$ cut in free inclination (blue circle centered at the forcing pole).
  Three TNOs with
  almost identical forcing centers
  are marked in blue crosses, with light blue being colds object ($I_\free < 4^\circ$) and dark blue being the hot object ($I_\free > 4^\circ$).
  If one use ecliptic inclinations to split the populations, $472231\ (2014\ \text{FU}_{71})$ would be misclassified as cold while $2015\ \text{GH}_{59}$
  (see Fig.~\ref{fig:two_tnos})
  would be misclassified as hot.
  TNOs in the overlapping area
  will maintain $I_\free<4^\circ$ and are correctly classified currently
  but some with $I_\free>2^\circ$, such as $2014\ \text{UY}_{228}$,  will
  cycle to ecliptic $I>4^\circ$ on Myr time scales.}
  \label{fig:free}
\end{figure*}

In Laplace-Lagrange secular theory, the barycentric inclination vector (also called the orbit pole) of a TNO (typically measured from the ecliptic plane) constantly rotates around its \textit{local forcing pole} under the perturbations of planets. When perturbed by a single planet, an object's forcing pole is constant and simply perpendicular to the planet's orbital plane. However, when perturbed by multiple planets, the local forcing pole constantly changes with time.
The time-dependent location of the forcing pole is predicted by the Laplace-Lagrange secular theory (Appendix~ \ref{appendix}), which gives the rectangular components of the forcing pole ($q=I\cos \Omega, p=I\sin \Omega$) at any given semimajor axis induced by the orbits of all giant planets.
\citet{Chiang.2008} showed that in the main classical belt, the forcing poles at various semimajor axes form a line in $(q, p)$ space, rotating around the location of the solar system's invariable pole with a 1.9 Myr period.
As the semimajor axis goes to infinity (although in practice needs only $a > 45$ au), the forcing pole approaches the invariable plane pole and is therefore fixed.

A TNO's current osculating orbital inclination is a sum of this locally forced inclination and its free inclination (also sometimes called the `proper' inclination).
By calculating and then subtracting the forced pole from an object's ecliptic inclination, the resultant free inclination vector components are obtained.
The magnitude of the \textbf{free inclination} vector is \textbf{$I_\free$}, and the phase provides the free ascending node $\Omega_\free$. In Fig.~\ref{fig:free} we illustrate the geometric relationship between the ecliptic (orange) inclination, the forced pole (red), and the free (blue) inclination in the rectangular $(q, p)$ space. 
We also recommend the non-expert reader to \citet{Gladman.2021}'s figure 1 and their supplemental video for more details.

\begin{figure}[htb!]
  \centering
  \includegraphics[width=1\columnwidth]{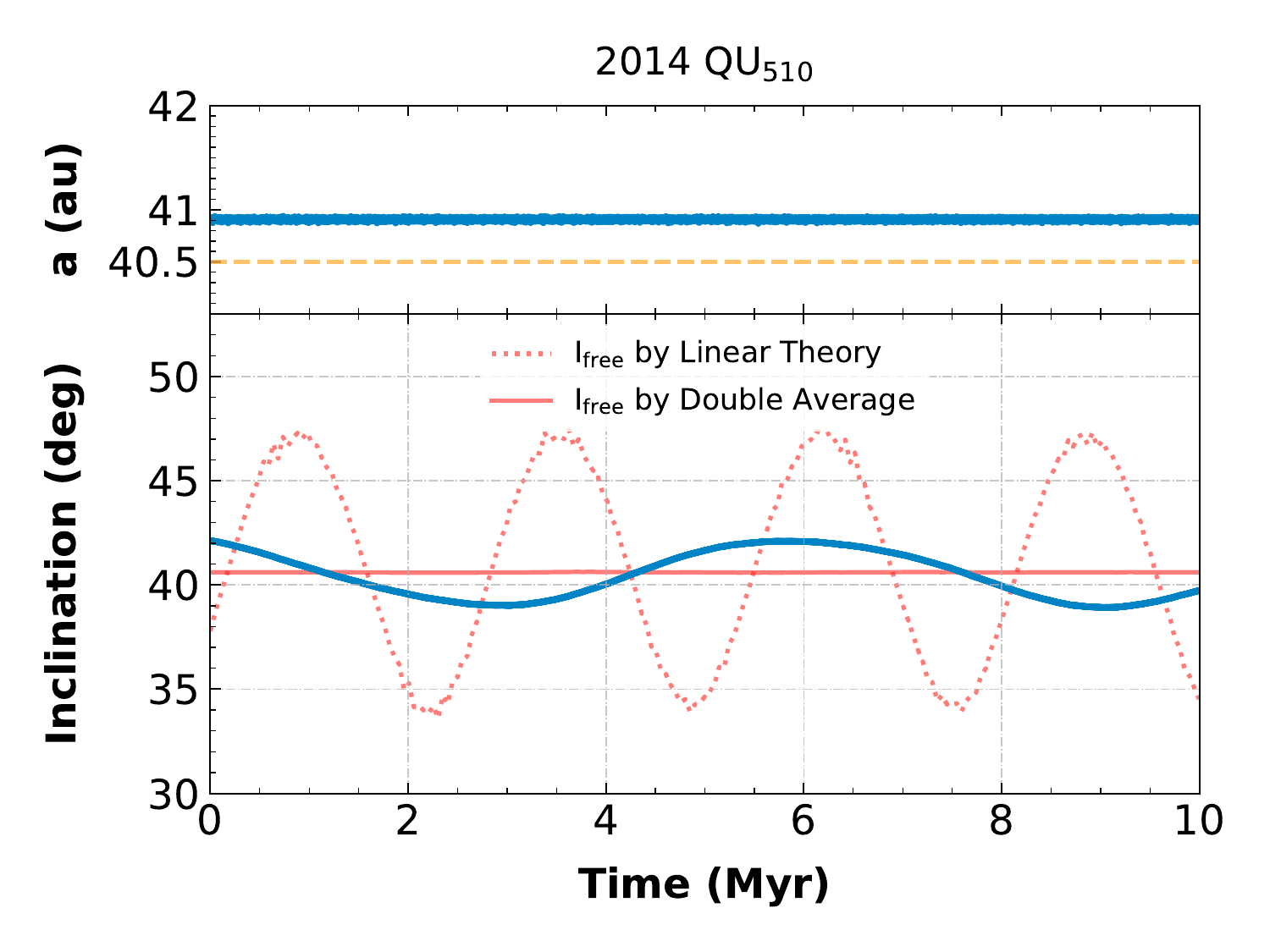}
  \caption{\textbf{Blue curves:} barycentric $a$ (top panel) and ecliptic $I$ (bottom panel) evolution of $2014\ \text{QU}_{510}$ over 10 Myr. The orange dashed line in the upper panel marks the semimajor axis of the $\nu_{18}$ secular resonance for circular and planar orbits.
  \textbf{Red curves:} $I_\free$ calculated by the linear theory (dotted curve) and the semi-analytical double average method (solid curve). The ranges of the blue curve, the dotted curve, and the solid line are 3.2$^\circ$, 14.4$^\circ$, and 0.06$^\circ$, respectively.; the double average method yields a much better-conserved value of $I_\free$. }
  \label{fig:near_secular_tno}
\end{figure}

In theory, the free inclination of a
non-resonant, non-scattering object is constant over time. The conservation of $I_\free$ at 38.6 au and at 43 au for 4 Gyr has been verified by numerical integrations \citep{Chiang.2008}. 
However, we find that near the $\nu_{18}$ secular resonance at $a=40.5$ au, the $I_\free$ of TNOs calculated by the linear theory are not conserved even over our much shorter 10 Myr integrations. Fig.~\ref{fig:near_secular_tno} shows the barycentric ecliptic $a$ and $I$ evolutions (blue curves) of $2014\ \text{QU}_{510}$ as well as its $I_\free$ calculated by the linear theory (red dotted curve).
To compute the linear secular $I_\free$ evolution of this object over our simulation, 
we recalculate the eigenmodes of the Solar System at each time output,
based on the constantly-evolving orbits of the 4 giant planets (using the method described in \citealt{Murray.1999}) and use those to determine $2014\ \text{QU}_{510}$'s forced plane and thus free inclination.
As shown in Fig.~\ref{fig:near_secular_tno}, the linear theory $I_\free$ of this object is not conserved at all; its amplitude even exceeds the variation of its osculating ecliptic inclination.

The varying $I_\free$ computed from linear theory near $a=40.5$ au demonstrates that the forcing pole is not correctly predicted near the secular resonance. The reason for this failure is that in the linear theory, the expected precession rate ($B$ in Appendix~\ref{appendix}) of a TNO is only a function of its semimajor axis. 
The real nodal precession rate, however, 
also depends on the object's eccentricity and inclination. Ignoring high-order terms in $e$ and $I$ produces an inaccurate precession rate, resulting in the incorrect determination of its forcing pole. This effect is particularly strong near a secular resonance, due to the fact that the term $B - f_j$ (where $f_j$ is an eigenfrequency) exists in the denominator of the forcing pole expression (Eq.~\ref{eq:forced_term}).

To get the correct $I_\free$, especially near the $\nu_{18}$, we adopt a semi-analytical method to recalculate the correct precession rate at every time step. It's based on numerically averaging the TNO's Hamiltonian over two `fast angles' (called the `double average' method hereafter; see 
\citealt{Morbidelli.2002}
for references). 
Simply put, the double average method (developed in the Appendix~\ref{sec:double_average})
calculates the precession rate as a function of $a$, $e$, $I$, and $\omega$, instead of just $a$ as in the linear theory. As a result, the method produces a rate closer to the TNO's true precession rate, and thus a more accurate forcing pole and a better conserved free inclination. Taking the object from Fig.~\ref{fig:near_secular_tno} as an example: the linear theory predicts a nodal rate of $-0.63''/\text{yr}$, which is very close to the $f_8 = -0.68''/\text{yr}$ inclination eigenfrequency of the solar system \citep{Brouwer.1950}. 
In contrast, the real precession rate according to numerical integration is only $-0.206''/\text{yr}$, which is much closer to what the double average method predicts (a rate varying between $-0.19''/\text{yr}$ to $-0.21''/\text{yr}$ as the orbit evolves). The $\sim40^\circ$ inclination makes the TNO precess slower than a planar orbit of the same $a$, pulling itself away from the $\nu_{18}$ secular resonance despite being near $a=40.5$ au. Obviously, the double average method provides a more accurate precession rate, resulting in a much better conserved $I_\free$ for the TNO (red solid line in Fig.~\ref{fig:near_secular_tno}). The details of the method are described in Appendix.~\ref{sec:double_average}.

\subsection{Results}\label{sec2.3}

\begin{figure*}[htb!]
  \centering
  \includegraphics[width=1.0\textwidth]{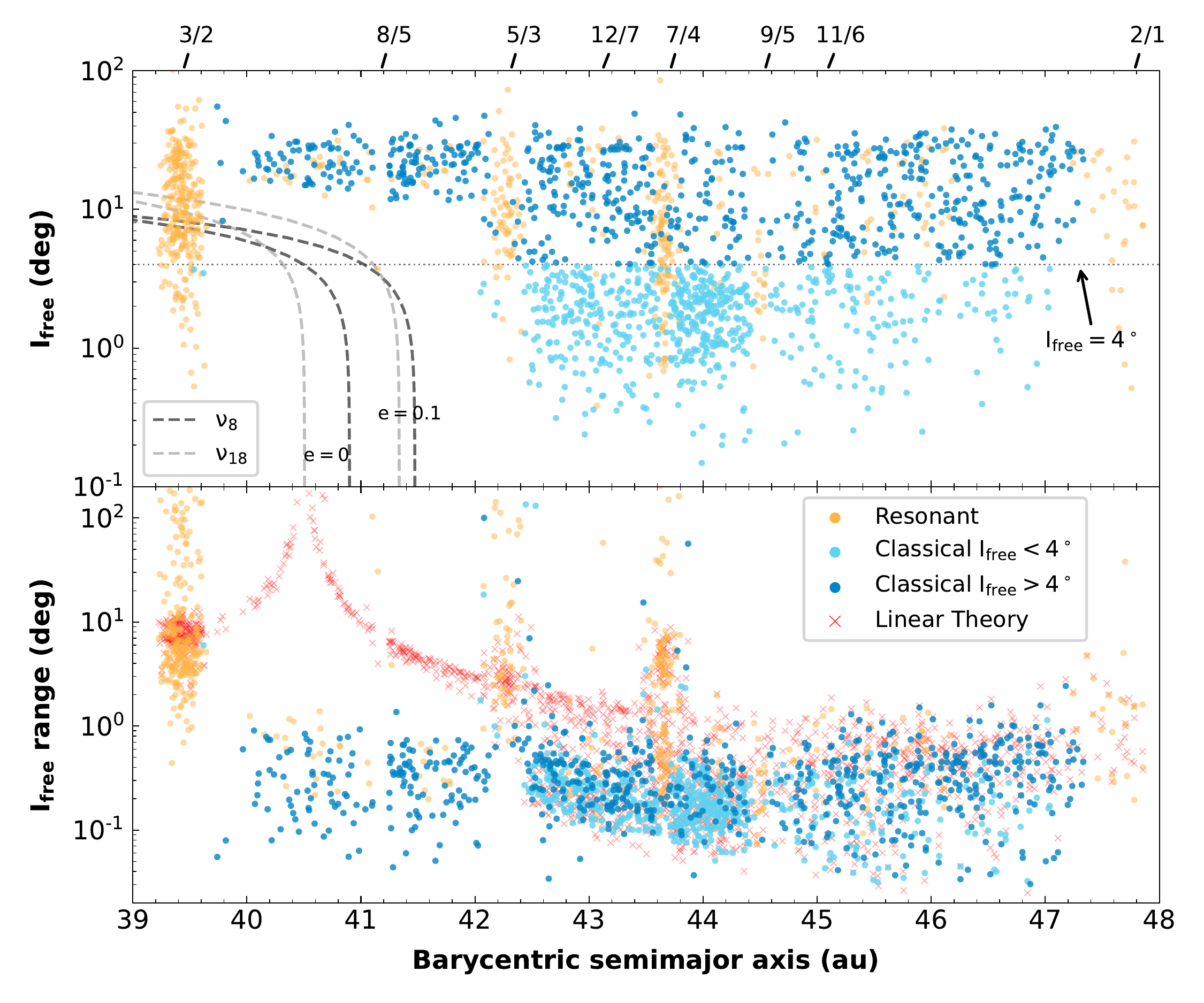}
  \caption{\textbf{Upper panel:} Our computed values of current $I_\free$ for each TNO as a function of its barycentric semimajor axis $a$ (note the log scale on the y-axis). Resonant objects are shown in orange and classical TNOs in blue, with light blue and dark blue denoting classical TNOs with $I_\free < 4^\circ$ and $I_\free > 4^\circ$, respectively.
  The dashed curves in the lower left portion of the plot give the center of the $\nu_{8}$ (black) and $\nu_{18}$ (gray) secular resonances, with curves on the left being $e=0$ and those on the right being $e=0.1$ (see text).
  \textbf{Bottom panel:} The range in calculated $I_\free$ values (log scale) over our 10 Myr integrations as a function of $a$.
  For objects with $a\simeq$ 39.7--42.5~au, the proximity of the $\nu_{18}$ secular resonance results in the TNO's expected precession rate being dramatically incorrect, resulting in widely-varying values of $I_\free$ when calculated using the linear theory (red crosses).
  When the nodal precession rate for non-resonant TNOs is corrected (using the double average method), well-preserved free inclinations are obtained for the classical TNOs (blue dots); resonant TNOs (orange dots) can have highly-variable values for their calculated $I_\free$ even with the double average method because the resonant dynamics are not accounted for.}
  \label{fig:result}
\end{figure*}

We applied both the linear theory and the double average method to the calculation of $I_\free$ for each of the non-scattering TNOs in our sample.
Fig.~\ref{fig:result}'s upper panel shows the double-averaging $I_\free$ as a function of $a$.
To illustrate the conservation of $I_\free$ over 10 Myr timescales, the bottom panel shows the variation, $I_\free$ range $\equiv$ $\text{max}(I_\free) - \text{min}(I_\free)$, over the integration for both methods.
The $I_\free$ range values show that, as expected, near the $\nu_{18}$ resonance at $40.5$ au, the linear theory fails to produce a well-conserved $I_\free$ (red crosses in Fig.~\ref{fig:result}), whereas even near the secular resonance our method
(blue dots) provides free inclinations that are as well conserved as for the rest of main-belt classicals. The vast majority of classicals have $I_\free$ conserved to better than $1^\circ$, although this is not the case for resonant objects (orange dots) and a handful of near-resonant objects.
Because the averaging method doesn't take into account the Hamiltonian's resonant terms, it cannot predict the correct nodal precession rate for objects affected by the mean motion commensurabilities. 
As a result, TNOs near and in the $3/2$ and the $5/3$ resonances have a significantly large $I_\free$ range (bottom panel of Fig.~\ref{fig:result}). 
The $7/4$ resonance, however, hosts TNOs with both large and small $I_\free$ ranges; 
the latter group are all objects with relatively small eccentricities ($e < 0.15$) and the vast majority have $I_\free < 10^\circ$.
Higher-order resonances in the main belt seem to have no clear effect on the object's $I_\free$ range, presumably due to their relatively weak strength.

Fig.~\ref{fig:result}'s lower panel aligns with our expectations: classical TNOs not affected by resonant dynamics generally have very small $I_\free$ variations, while resonant TNOs (especially those in strong, low-order resonances) have significantly large $I_\free$ ranges.
In other words, in Fig.~\ref{fig:result}'s upper panel, only for the classical objects (blue dots) can $I_\free$ be trusted to 
be cosmogonically relevant.

\begin{figure*}[htb!]
  \centering
  \includegraphics[width=0.9\textwidth]{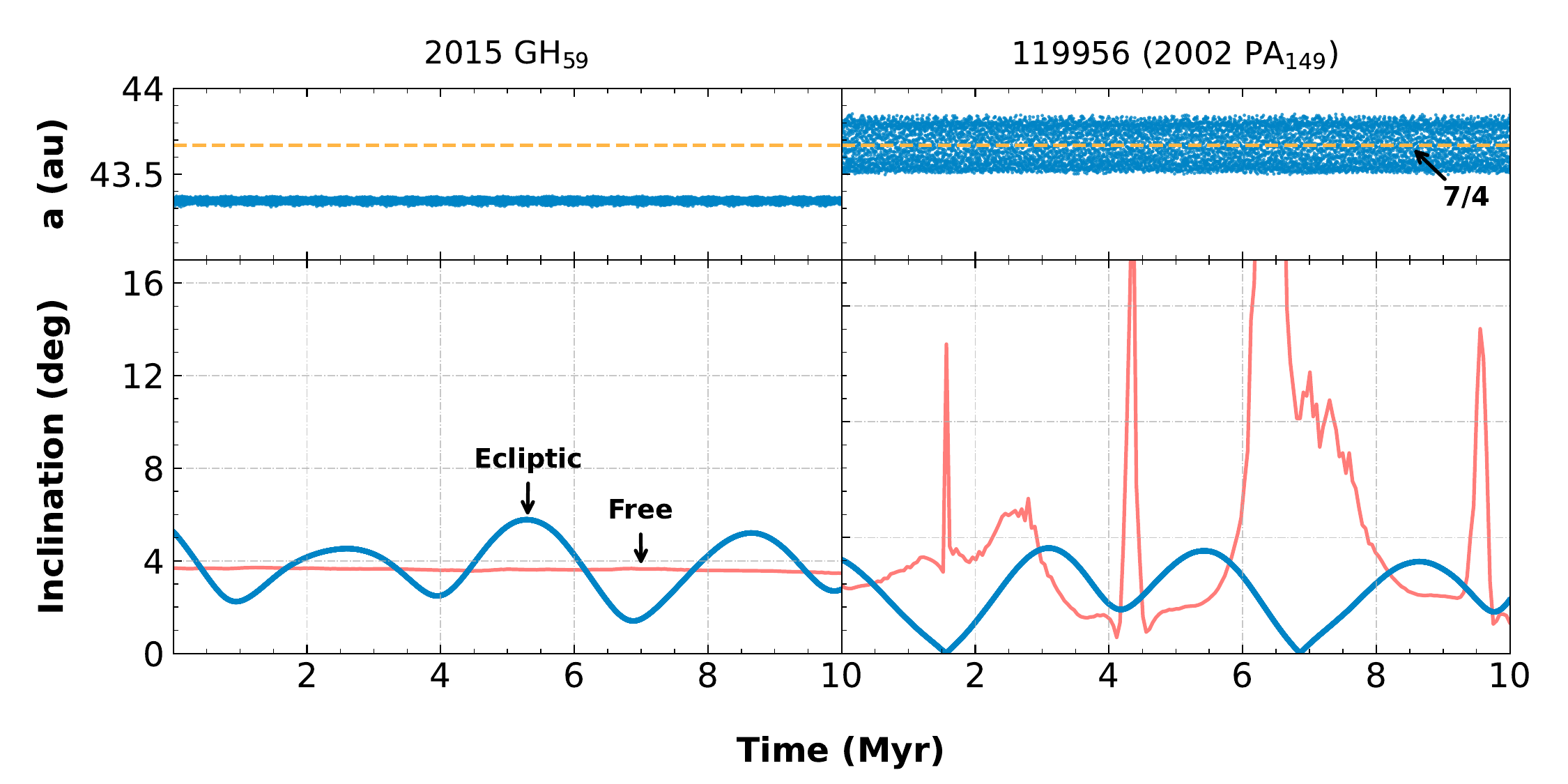}
  \caption{Evolution in $a$ (upper panels; blue curves), ecliptic $I$ (lower panels; blue curves) and $I_\free$ (lower panels; red curves) over 10 Myr for classical TNO 2015 GH$_\text{59}$ (left) and resonant TNO 119956 (2002 PA$_\text{149}$; right).
  The classical TNO has initial J2000 $I > 5^\circ$, but its $I_\free$ is always less than $4^\circ$ (see also Fig.~\ref{fig:free}'s right panel).
  The resonant object is in the $7/4$ resonance and its $I_\free$ (as calculated by the double average method) is not conserved; this is expected because neither linear secular theory nor the double average method is appropriate for resonant TNOs.}
  \label{fig:two_tnos}
\end{figure*}

If we limit our scope only to the classical TNOs in the $I_\free - a$ distribution, there are a few things worth pointing out: 

(1) Almost every classical TNO between the 3/2 and the 5/3 neptunian resonances has $I_\free$ larger than $10^\circ$, due to the presence of $\nu_8$ and $\nu_{18}$ secular resonances in the low-$I$ region. 
The $\nu_8$ resonance will excite eccentricities for lower-$I$ orbits to Neptune crossing, resulting in TNO
removal.
We computed the positions of these resonances (gray and black dashed curves) by iteratively converging (for a given $e$) to the resonant secular frequency ($g_8$ or $f_8$) 
by varying $I_\free$; the $e=0.1$ curve is very similar to those shown
in \citet{Knezevic.1991} and \citet{Morbidelli.2002}.

(2) The 8/5 resonance might be viewed as surprisingly devoid of TNOs, leaving a semimajor axis gap in the hot population  at $a$ = 41.2~au. 
We integrated five of the 8/5 resonant objects to 4 Gyr; none of them survive for the age of the Solar System, with a median dynamical lifetime of only 700 Myr. 
None of the real objects are thus deeply embedded and stabilized by the
mean-motion resonance for the age of the Solar System.
Given that the nearby third-order 7/4 resonance does not deplete on this
same time scale (even at large $I_\free$), this contrast seems
puzzling.
It is plausible that the proximity to the two secular resonances is
contributing to this instability, but may instead imply
something 
about capture into this resonance out of the abundant ancient scattering
population.

(3) There is a few-degree wide sparsely populated region
in the free inclination distribution (sitting just above $I_\free = 4^\circ$)
in the semimajor axis range between the 5/3 and the 7/4 resonances in the main belt.
This `gap' is nearly devoid of TNOs and exists only in the $I_\free$ space; it would be completely hidden if one were to plot the distribution using ecliptic $I$.
This explains why \citet{Laerhoven.2019} found isolating
$I_\free < 4^\circ$ is an excellent way to reduce 
contamination between hot and cold, minimizing interlopers when measuring the width of the cold population's 
inclination distribution.
We expand upon this in the Discussion section.

To show how the resonant dynamics affects $I_\free$, we plot 
(Fig.~\ref{fig:two_tnos})
the orbital evolution of
an object (119956 = 2002 PA$_\text{149}$)
trapped in the $7/4$ resonance with moderate $e\simeq$ 0.17
and 
a nearby cold classical TNO (2015 GH$_\text{59}$).
 Despite maintaining a relatively low ecliptic inclination, the resonant object's calculated free inclination is extremely variable (right panel of Fig.~\ref{fig:two_tnos}) because the assumptions underlying the linear secular or the double average $I_\free$ calculation are not valid for resonant objects; this highlights why we needed to classify our TNO sample prior to determining free inclinations. In contrast, the classical TNO in the left panel of Fig.~\ref{fig:two_tnos} demonstrates
why $I_\free$ is superior to the ecliptic inclination in separating objects into cold versus hot populations: 2015 GH$_\text{59}$ is an object with a $5.3^\circ$ ecliptic inclination currently, which would place it in the hot population according to most ecliptic inclination cuts. In addition, the 10-Myr average of its ecliptic inclination is $4.4^\circ$, still above a typical $4^\circ$ cut. However, our calculation shows its free inclination is only $3.69^\circ $ with a range of $0.25^\circ$, which keeps it always below a $4^\circ$ cut and thus always a cold object.

\begin{figure}[htb!]
  \centering
  \includegraphics[width=1\columnwidth]{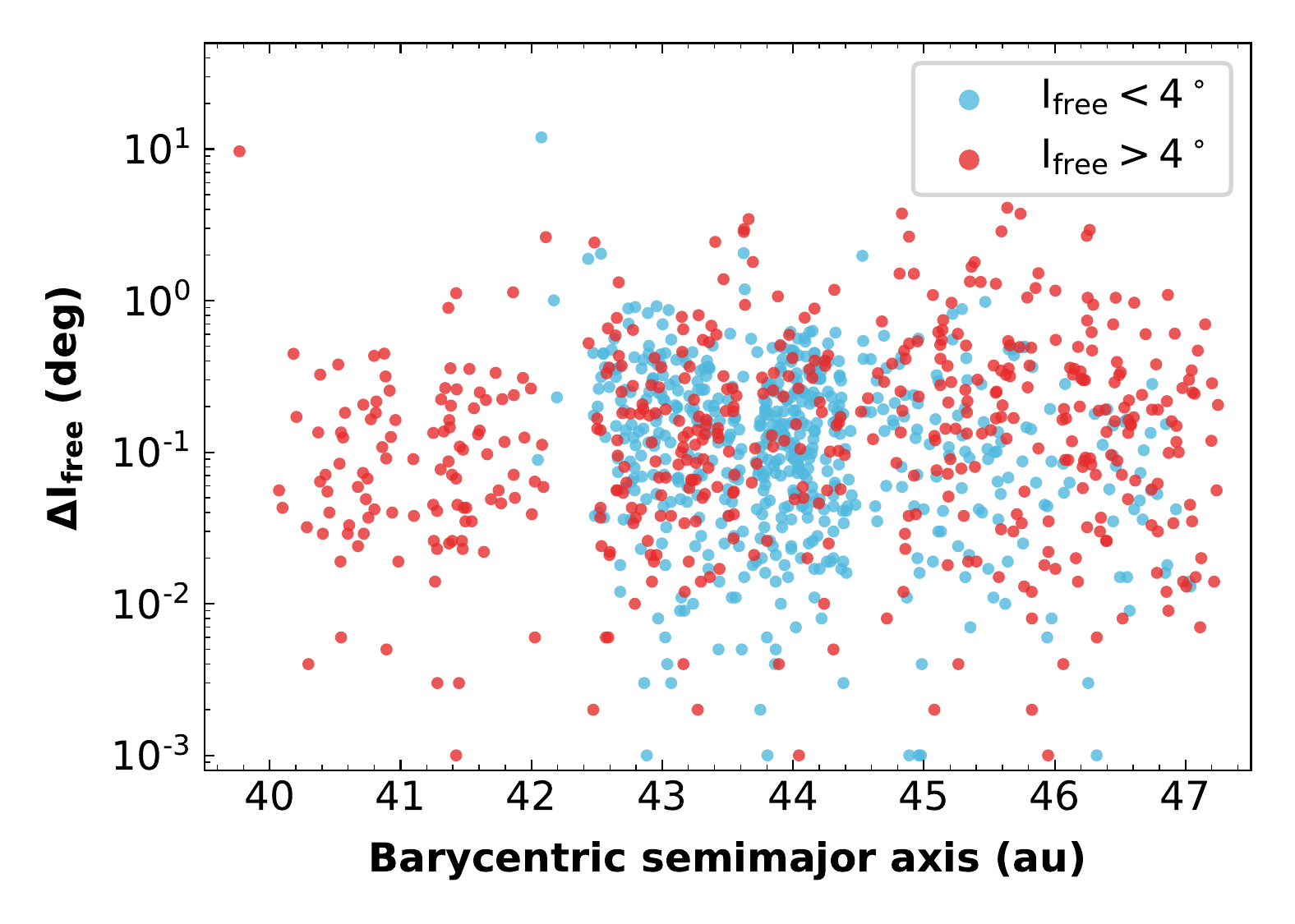}
  \caption{$\Delta I_\free$ measured across a 4 Gyr integration as a function of barycentric $a$ for main-belt classical TNOs.
  The blue and red dots represent $I_\free < 4 ^\circ$ and $I_\free > 4 ^\circ$, respectively.
  The majority of TNOs in both categories have values of $I_\free$ that are conserved to within $1^\circ$ even over 4 Gyr.}
  \label{fig:4gyr}
  \end{figure}

We demonstrated above the conservation of $I_\free$ for classical objects over 10 Myr timescales. But what one truly cares about is whether $I_\free$ is stable for the age of the Solar System.
In other words, will the current $I_\free$ of classical TNOs reflect their $I_\free$ $\sim$4 billion years ago, at end of the giant planet formation and migration/transport to their present-day orbits?
To answer this, we extended the numerical integrations of the observed objects for 4 Gyr and plotted the surviving particles' $\Delta I_\free$ (the absolute difference between each object's current $I_\free$ and that at 4 Gyr) in Fig.~\ref{fig:4gyr}.
The vast majority of classical TNOs, no matter what their current inclinations are, have $\Delta I_\free < 1 ^\circ$.
The very few outliers are mainly distributed around major mean-motion resonances, indicating some occasional interactions with the resonances during the 4 Gyr evolution in which the secular conservation is lost. Fig.~\ref{fig:4gyr} thus shows that the $I_\free$ distribution we compute today for the classical TNOs can be taken to be representative of the primordial distribution.

Last but not least, we also explored the idea of whether the (easily calculated) mean ecliptic inclination from numerical integrations can serve as a good proxy for $I_\free$.
We find that for classical TNOs whose $I_\free > 4^\circ$, the mean ecliptic inclination over 10 Myr is a good approximation to $I_\free$, with the median absolute difference being only $0.2^\circ$.
However, in the cold population ($I_\free < 4^\circ$), the median absolute difference is $0.6^\circ$, which renders the mean ecliptic inclination a low-quality estimate of the free inclination for this
population (see the example discussed above from Fig.~\ref{fig:two_tnos}).
Moreover, we point out if one intended to study the main belt's inclination distribution through the mean ecliptic inclination, 
this averaging would result in all objects with $I_\free$ smaller than the local forced inclination being assigned a mean ecliptic inclination of roughly the forced value of about $2^\circ$;
the distribution of very low $I_\free$ objects would be completely erased.
These same arguments apply if one tried to use the invariable plane
as the reference (rather than the ecliptic); the cold population's
median absolute inclination difference is still $0.5^\circ$.
It is thus superior to use $I_\free$ computed by the double average to studying the main belt's inclination distribution.

\section{Discussion}\label{discussion}

As a summary, the free inclination distribution we have computed 
for the main belt 
(Fig.~\ref{fig:result}) 
illustrates several points:

1. The innermost boundary of the cold population at $a \simeq$ 42~au
is being set by the existence of the secular resonances.
The absence of low-$I$ TNOs here does not imply that the cold
belt did not exist here before the giant planets finished formation 
and migration.
How and when these secular resonances reached their current location
is a subject of much speculation
(egs., \citealt{10.1088/0004-637x/738/1/13}, 
\citealt{Dawson.2012},
\citealt{Gladman.2012},
\citealt{Nesvorny.2018},
\citealt{10.1016/j.icarus.2019.113417}, and references therein)

2.  The double averaging method removes the apparent forced-inclination singularities that occur in the linear secular theory.
If one wished to study the secular effects of additional 
planets on the ancient or current structure of the Solar System
\citep[egs,][]{Volk.2017, Batygin.2019}, this method is to be preferred
to estimate the inclination perturbations produced by the planet.

3.  The preservation of the $I_\free$ calculated via this method over
4 Gyr allows us to study the ancient inclination structure of
the belt (that is, the structure existing at the end of the the
planet formation epoch).  
This reinforces the idea that there was a cold ($I_\free<4^\circ$)
population present at that time which (at least in the 
$a<44.5$~au region of the main belt) is well separated from the
presumably implanted hot population.

\begin{figure}[htb!]
  \centering
  \includegraphics[width=1\columnwidth]{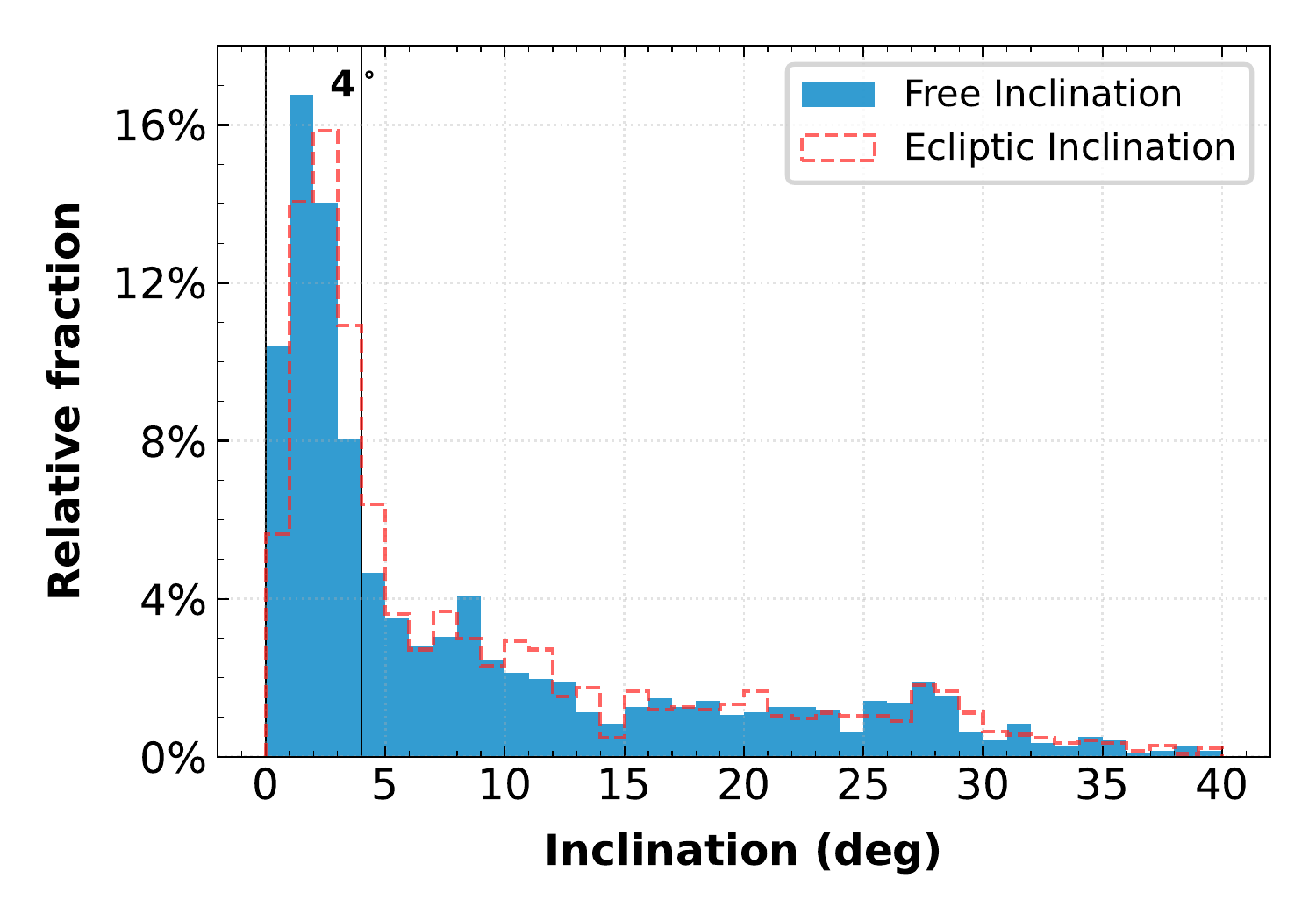}
  \caption{A fractional histogram of free inclinations (blue bars) and 
  ecliptic inclinations (red outlines) for 1450 main-belt classical TNOs beyond the 
  5/3 resonance ($a >$ 42.3 au).   Only a few such objects (not shown) have $>40^\circ$ inclinations.
  Note that because most TNO surveys have been conducted at relatively low latitudes, there is an observational bias against the large-inclination TNOs; thus the high-$I$ tail in this histogram should not be taken to represent the true hot TNO inclination distribution.
  However, the dramatic spike at low
  inclinations (which is due to the cold population) is still clear in the observed population, and even more obvious in
  the $I_\free$ histogram (which is shifted towards zero due to the true secular dynamics
  of the belt).
  A 4-degree cut is shown by the vertical line; few TNOs in the 0-4$^\circ$ range 
  would be hot-population interlopers.}
  \label{fig:hist}
  \end{figure}

4.  
The main-belt TNO population
clearly has multiple superposed components 
\citep[see][and citations to it]{Brown.2001} and,
because there 
is evidence that these components have
different
physical properties due to different formation locations, 
there needs to be some way to 
easily
separate observed TNOs
for spectrophotometric studies.
Due to the narrow width of the cold component, TNOs with large ecliptic
inclinations (larger than 10$^\circ$, say) will almost all be from
the hot component with very few interlopers.
At small inclinations, the majority of
the TNOs will be from the cold population, with the 
interloper fraction depending on the component
$I$~distributions and relative populations.
The often-used 5$^\circ$ cut in ecliptic inclination 
to separate the cold and hot components
was proposed by
\cite{Bernstein.2004}, although no justification for that particular
value was given.

Here we have demonstrated the superiority of using a cut in $I_\free$ to isolate the hot and cold populations.
\citet{Laerhoven.2019} found that the cold population is tightly confined  (in $I_\free$)
around the local forcing pole; this is especially true in the inner part of the main belt, where
they found the cold population has an inclination width of just  $\simeq1.8^\circ$.
Fig.~\ref{fig:hist} shows histograms of ecliptic and free inclination for the non-resonant main belt TNOs in this work. 
The low-$I$ peak in the free inclination histogram is sharper than in the ecliptic histogram, and there is a noticeable drop-off in the observed population at $I_\free=4^\circ$ (roughly twice the cold population's inclination width).
We note that observational biases are not accounted for in Fig.~\ref{fig:hist}, though they were accounted for in \citet{Laerhoven.2019}'s 
analysis of the cold population's inclination width.
Based on that inclination width and our analysis here, we suggest that $I_\free<4^\circ$ is a reasonable choice when using a simple cut to separate the hot and cold populations in the main TNO belt.

\section{Data Release}\label{sec3}

\begin{table*}[htb!]
  \caption{Barycentric elements and $I_\free$ for main-belt TNOs}
  \begin{center}
  \tabcolsep5pt
  \begin{tabularx}{\textwidth}{p{0.28\textwidth}p{0.08\textwidth}p{0.59\textwidth}}
  \hline\hline
  Column names & Units & Descriptions \\
  \hline
  $a$ ({\tt a}) & au & Semimajor axis of the nominal orbit \\
  $e$ ({\tt e}) &   & Eccentricity \\
  $I$ ({\tt inc}) &  deg & Inclination \\
  $\Omega$ ({\tt Omega}) & deg & Longitude of ascending node \\
  $\omega$ ({\tt omega}) & deg & Argument of perihelion \\
  $M$ ({\tt M}) & deg & Mean anomaly \\
  $d$ ({\tt dist}) & au & Distance from the barycenter \\
  $H$ ({\tt H}) & mag & Absolute magnitude \\
  {\tt RESO} &  & Dynamics flag: -1 for scattering, 0 for non-resonant/classical, $>0$ for the exactly resonant ratio (e.g. `74' stands for the 7/4 mean-motion resonance with Neptune) \\
  $I_\free$ ({\tt Ifree}) & deg & Free inclination, computed via double averaging \\
  $I_\free$ range ({\tt IfreeRange}) & deg & Free inclination range over 10-Myr integration time \\
  $q_\text{forced}$ ({\tt qForced}) & deg & q component of the forcing pole. $q_\text{forced} = I_\forced \cos{(\Omega_\text{forced})}$ \\
  $p_\text{forced}$ ({\tt pForced}) & deg & p component of the forcing pole. $p_\text{forced} = I_\forced \sin{(\Omega_\text{forced})}$ \\
  {\tt OSSOS} &  & OSSOS++ internal designation (`x' for non-OSSOS++ objects) \\
  {\tt DES} &  & DES internal designation (`x' for non-DES objects)  \\
  {\tt ID} &  & ID downloaded from JPL Small-Body Database: For a numbered TNO, ID gives its designated number; For an unnumbered TNO, ID gives its compact provisional designation   \\
  {\tt Name} &  & Full name (designation in bracket)  \\
  \hline
  \end{tabularx}
  \end{center}
  {\bf Note.} The first six orbital elements and the distance are barycentric and in the IAU76/J2000 ecliptic reference frame, referring to epoch JD 2459400.5. Both $I_\free$ and $I_\free$ range are independent of the choice of reference frame. The absolute magnitude $H$, id and full name are directly retrieved from JPL on October 5th, 2021; these values  could change as the MPC receives additional observations. {\bf This table is available as a downloadable, machine readable file \footnote[1]{Download this table on \url{https://yukunhuang.com}. The data file will also be available on the journal website when published.}.}
  \label{tab:free}
\end{table*}

The purpose of this work is to provide tabulated barycentric orbit elements and $I_\free$ for the currently observed main classical belt TNOs as a resource for studies comparing the hot and cold populations.
We do this in Table~\ref{tab:free}.
We identify each TNO in our sample by their primary MPC designation, but we also include OSSOS++ and DES designations for objects that appear in either survey so that one could more easily use the survey simulators
of those surveys for quantitative debiasing.
The absolute magnitude $H$ in Table~\ref{tab:free} is taken from the JPL Small body database; we note that these $H$ values use approximate color transformations and should not be used with the survey simulators mentioned above as they are not linked to a specific filter.

In addition, to help the reader quickly estimate the correct $I_\free$ for future TNOs, we provide Table~\ref{tab:forced}, in which the forcing pole components $(q, p)$ are pre-computed in a $(a, e, I, \omega)$ 4-dimensional grid.
We also provide a Python script to read the file and find the closest data point for any given orbit, which the reader can then use to estimate the $I_\free$ that would be given by the double average method.
However, it's important to note that this simplified approach of evaluating $I_\free$ can only be trusted if the TNO: (1) is a non-resonant and non-scattering object within the given orbital ranges, (2) stays away from the $\nu_{18}$ secular resonance (in other words, the forcing pole is relatively small), and (3) has a current inclination computed at the current epoch. We have tested the file and confirmed that for TNOs that meet these three requirements, this script yields $I_\free$  to a precision of $\sim0.1^\circ$ compared to that computed by double average.

\begin{table}[htb!]
  \caption{$q, p$ components of the forcing pole for various orbital elements}
  \begin{center}
  \tabcolsep5pt
  \begin{tabularx}{1.0\columnwidth}{p{0.2\columnwidth}p{0.25\columnwidth}p{0.18\columnwidth}p{0.13\columnwidth}}
  \hline\hline
  Axis names & Range & Grid sizes & Dimensions \\
  \hline
  $a$ & (39.4, 47.7) au & 0.1 au & 84 \\
  $e$ & (0, 0.25) & 0.01 & 26 \\
  $I$ & (0, 40) deg & 2 deg & 21 \\
  $\omega$ & (0, 90) deg & 10 deg & 10 \\
  \hline
  \textbf{Total Size} &  &  & 458,640 \\
  \hline
  \end{tabularx}
  \end{center}
  {\bf Note.} In the double average method, the forcing pole vector is a function of $(a,e,I,\omega)$. For each data point on grid, we gives the $q, p$ components of the forcing pole in degrees, which can be used to estimate the $I_\free$ of nearby orbits. $\omega$ is only in the range of $(0, 90)\ \text{deg}$ due to its two-fold reflection symmetries in both the orbital plane and the central axis.
  \label{tab:forced}
\end{table}

With accurate values of $I_\free$ and the knowledge of a cleaner separation between hot and cold population with a $4^\circ$ boundary in this variable, one can use our tabulated $I_\free$ values to reduce the occurrence of cross-contamination between the two groups in photometric and spectroscopic studies of those populations whose goal is to constrain primordial TNO surface properties.  
Lastly, we provide a rapid method allowing anyone to easily estimate the correct
$I_\free$ for future TNO discoveries in the main Kuiper Belt.

\section{Acknowledgements}

We thank J.-M. Petit for valuable discussions, P. H. Bernardinelli for providing us the DES Y6 data ,and an anonymous referee for helpful improvements. YH acknowledges support from China Scholarship Council (grant 201906210046) and the Edwin S.H. Leong
International Leadership Fund, BG acknowledges Canadian funding support from NSERC, and KV acknowledges support from NSF (grant AST-1824869) and NASA (grants 80NSSC19K0785 and 80NSSC21K0376).
This work used High Performance Computing (HPC) resources supported by the University of Arizona TRIF, UITS, and Research, Innovation, and Impact (RII) and maintained by the UArizona Research Technologies department.

\pagebreak

\bibliography{ref}{}
\bibliographystyle{aasjournal}

\appendix

\section{Free Inclination Algorithm}\label{appendix}
\subsection{Laplace-Lagrange Secular Theory}\label{sec:linear_thoery}
The motions of the planets in the Solar System is a non-integrable N-Body problem. With suitable assumptions, it is possible to express the long-term variations of the orbits of the solar system bodies in an analytical form. A widely used solution is derived by \citet{Brouwer.1950}, where 10 frequencies for the $e-\varpi$ solution and 8 frequencies for the $I-\Omega$ solution are given. We denote them the eigenfrequencies $g_1$ -- $g_{10}$ and $f_1$ -- $f_8$ of the solar system. Here, we recap the secular perturbation theory for test particles and give the explicit equations for calculating their $I_\free$ (see section 7 of \citealt{Murray.1999}).

 The orbital variations of TNOs are strongly influenced by perturbations from the planets. To study this, we start by transforming the orbital elements to the coordinates:
\begin{equation}\label{eq:qp_cords}
    q = I \cos \Omega, \quad p = I \sin \Omega,
\end{equation}
where $I$ is the osculating orbital inclination and $\Omega$ is the osculating longitude of ascending node in some chosen reference frame. With the computed eigenfrequencies and eigenmodes of the planetary motions, it is possible to write down the solutions for small body inclinations in the new $(q, p)$ coordinates:
\begin{equation}\label{eq:lapalce-lagrange_secular_solution}
    q = I_\free \cos\left(B t+\gamma\right) + q_\forced(t), \quad
    p = I_\free \sin\left(B t+\gamma\right) + p_\forced(t),
\end{equation}
where $I_\text{free}$ is the free inclination, and $q_\forced$ and $p_\forced$ are components of the forcing poles imposed by planetary perturbations. In the present work, only the four giant planets are taken into account, so the resulting forced terms are given by
\begin{equation}\label{eq:forced_term}
    q_\forced(t) = - \sum_{j = 5}^{8} \frac{\mu_j}{B-f_j} \cos(f_j t + \gamma_j), \quad
    p_\forced(t) = - \sum_{j = 5}^{8} \frac{\mu_j}{B-f_j} \sin(f_j t + \gamma_j),
\end{equation}
where $j$ denotes the index of the inclination eigenfrequencies/eigenvectors ($f_j$ and $I_{ji}$ below) of the Solar System, and $B$ denotes the expected precession rate of the small body's node. 
In Laplace-Lagrange secular theory, $B$ is the summation of precession rates contributed by each planet ($B_j$), which depend on both the planetary and the small body's semimajor axes ($a_j$ and $a$). 
Note that the term $B-f_j$ appears in the denominators of both equations; when the expected precession rate $B$ approaches any Solar System eigenfrequency $f_j$ (or $g_j$ for the eccentricity frequency), the forced vector would diverge, which corresponds to the secular resonance. The inclination secular resonance sends the TNO into a large-$I$ oscillation, while the eccentricity one will destabilize the object by boosting its orbital eccentricity to planet-crossing values.

For TNOs in the main belt (where $\alpha_j = a_j/a < 1$ always holds), $B$, $B_j$, and $\mu_j$ are given by:
\begin{equation}\label{eq:B_Bj_mu}
  \begin{aligned}
    B_j &= \frac{1}{4} \frac{m_j}{m_\odot} n \alpha_j  b_{3/2}^{(1)}(\alpha_j), \\
    B &=  - \sum_{j = 5}^{8} B_j, \\
    \mu_j & = \sum_{j = 5}^{8} B_j I_{ji},
  \end{aligned}
\end{equation}
where $m_\odot$ is the solar mass, $m_j$ is the mass of the $j$-th planet, and $n$ is the mean motion of the small body. $b_{3/2}^{(1)}(\alpha_j)$ is the Laplace coefficient.

The Laplace-Lagrange secular theory predicts a nodal precession rate for each TNO, under the assumption that both the planets and the TNOs have near circular and planar orbits. As a result, the forced inclination $I_\text{forced}$ and the longitude of ascending node $\Omega_\text{forced}$ are both functions of the semimajor axis only.

\subsection{Free Inclination with Double Average}\label{sec:double_average}
As shown in the main text, the linear theory gives erroneous free inclinations for TNOs near the $\nu_{18}$ secular resonance, and poorly-conserved free inclinations for moderately-inclined ($I \sim 20^\circ$) objects in the main classical belt. We therefore provide a more accurate way to calculate the expected precession rate $B$ of a small body. This calculation is performed every output time step of the numerical integration, resulting a better-measured forcing pole and thus a better-conserved $I_\free$.

Our method is based on numerically averaging the leading Hamiltonian over the two most quickly varying angles (hereafter called the `double average method'), which avoids any truncation in powers of the small body's eccentricity and inclination\citep{Williams.1969}. 
This approach has been commonly used in Solar System studies \citep{Henrard.1990, Morbidelli.1991, Froeschle.1994, Michel.1997}, in order to compute the locations of secular resonances over a large range of $e$ and $I$. We summarize the major steps to obtain the correct precession rates ($B$ and $B_j$) using the double average method (see \citealt{Michel.1997} and chapter 8 of \citealt{Morbidelli.2002} for more complete details).

One first introduces the canonical Delaunay variables:
\begin{equation}\label{eq:delaunay_variables}
\begin{array}{ll}
L=\sqrt{a} &\quad  l=M \\
G=\sqrt{a\left(1-e^{2}\right)} &\quad  g=\omega \\
H=\sqrt{a\left(1-e^{2}\right)} \cos I &\quad  h=\Omega,
\end{array}
\end{equation}
where the semimajor axis $a$, the eccentricity $e$, the inclination $I$, the argument of perihelion $\omega$, the longitude of the ascending node $\Omega$, and the mean anomaly $M$ are the usual Keplerian orbital elements.

Assuming all the planets to be on co-planar circular orbits, the Hamiltonian of a TNO perturbed by the $j$-th planet can be written as
\begin{equation}\label{eq:Hamiltonian}
  \mathcal{H}= \underbrace{-\frac{1}{2 L^{2}} \vphantom{-\sum_{j=5}^{8}}}_{\mathcal{H}_\text{kep}} +\underbrace{\vphantom{-\sum_{j=5}^{8}}m_{j} \mathcal{P}_{j}\left(L, G, H, L_{j} ; l, g, h, l_{j}\right)}_{\mathcal{H}_\text{sec}^{(j)}},
\end{equation}
where $\mathcal{H}_\text{kep}$ is the integrable Keplerian motion of the TNO around the Sun and $\mathcal{H}_\text{sec}^{(j)}$ accounts for the planetary perturbation by the $j$-th planet, in which $\mathcal{P}_{j}$ is the normalized term that only depends on orbital elements. Assuming the TNO is not trapped inside a mean-motion resonance, then the secular Hamiltonian $\mathcal{H}_\text{sec}^{(j)}$ can be averaged over the two unrelated fast angles, $l$ and $l_j$ (a `double average'), which yields
\begin{equation}\label{eq:average_Hamiltonian}
  \overline{\mathcal{H}}_\text{sec}^{(j)}= -m_{j} \mathcal{P}_{j}\left( G, H ; g, h\right),
\end{equation}
To write the averaged Hamiltonian in explicit form:
\begin{equation}\label{eq:Hamiltonian_sec_double}
  \overline{\mathcal{H}}_\text{sec}^{(j)}=-\frac{\mathcal{G}}{(2 \pi)^{2}}  \int_{0}^{2 \pi} \int_{0}^{2 \pi}\left(\frac{1}{\left\|\boldsymbol{\Delta}_{j}\right\|}-\frac{\mathbf{r} \cdot \mathbf{s}_{j}}{\left\|\mathbf{s}_{j}\right\|^{3}}\right) \mathrm{d} l \mathrm{~d} l_{j} ,
\end{equation}
in which the vectors $\mathbf{r}$ and $\mathbf{s}_{j}$ denote the heliocentric positions of the small body and of the $j$-th planet, respectively, and $\boldsymbol{\Delta}_{j} = \mathbf{r} - \mathbf{s}_{j}$. Under the approximation that eccentricities and inclinations of the planets are zero, one of the integrals can be analytically computed using the complete elliptic function of the first kind $K$:
\begin{equation}\label{eq:ring_potential}
   \int_{0}^{2 \pi}\left(\frac{1}{\left\|\boldsymbol{\Delta}_{j}\right\|}-\frac{\mathbf{r} \cdot \mathbf{s}_{j}}{\left\|\mathbf{s}_{j}\right\|^{3}}\right) \mathrm{d} l_{j}=\int_{0}^{2 \pi} \frac{1}{\left\|\boldsymbol{\Delta}_{j}\right\|} \mathrm{d} l_{j}=\frac{4}{\sqrt{r^{2}+a_{j}^{2}}} \sqrt{1-\frac{\mu}{2}} K(\mu),
\end{equation}
where
\begin{equation}\label{eq:m}
  \mu=\frac{4 a_{j} \sqrt{x^{2}+y^{2}}}{r^{2}+a_{j}^{2}+2 a_{j} \sqrt{x^{2}+y^{2}}},
\end{equation}
and $x$ and $y$ are the coordinates of $\mathbf{r}$'s projection on the plane of the planetary orbit, with $r=\left\|\mathbf{r}\right\|$.
It is worth noting that Eq.~\ref{eq:ring_potential} has the physical interpretation of the potential from a homogeneous ring (averaging the Hamiltonian over a planet's mean anomaly $l_j$ is equivalent to spreading out the planetary mass on a circular ring of radius $a_j$).
Combining Eq.~\ref{eq:Hamiltonian_sec_double} through \ref{eq:m}, we have
\begin{equation}\label{eq:H_j}
  \begin{aligned}
  \overline{\mathcal{H}}_\text{sec}^{(j)} &= -\frac{\mathcal{G}}{\pi^{2}} m_j \int_{0}^{2 \pi} \sqrt{\frac{1-\mu/2}{r^{2}+a_{j}^{2}}}  K(\mu) \mathrm{d} l, \\
  \overline{\mathcal{H}}_\text{sec} &= \sum_{j=5}^{8} \overline{\mathcal{H}}_\text{sec}^{(j)},
  \end{aligned}
\end{equation}
where $\overline{\mathcal{H}}_\text{sec}$ is the Hamiltonian accounts for the total planetary perturbations. The expected nodal precession rate contributed by each planet $B_j$ and the total precession rate $B$ can thus be obtained through numerical differentiation
\begin{equation}\label{eq:B_j}
  \begin{aligned}
  B_j &= \dot{h}^{(j)} = \frac{\partial \overline{\mathcal{H}}_\text{sec}^{(j)}}{\partial H}, \\
  B &= - \sum_{j = 5}^{8} B_j,
  \end{aligned}
\end{equation}
and the resulted $B_j$ and $B$ are not only functions of $a$ and $a_j$, but also functions of $e,I$, and $\omega$. Replacing the nodal precession rates from Eq.~\ref{eq:lapalce-lagrange_secular_solution} to \ref{eq:B_Bj_mu}, we get the correct forcing pole and thus the correct free inclination for each TNO.
\end{document}